\documentclass[12pt,a4paper]{article}
\usepackage{epsfig}
\makeatletter

\def\vereq#1#2{\lower3pt\vbox{\baselineskip0.5pt\lineskip0.5pt
\ialign{$\m@th#1\hfill##\hfil$\crcr#2\crcr\sim\crcr}}}
\def\gtrsim{\mathrel{\mathpalette\vereq>}}
\makeatother

\newcommand{\astar}{\ensuremath{a_*}}

\newcommand{\rstar}{\ensuremath{r_{*}}}

\newcommand{\omtil}{\ensuremath{\widetilde{\omega}}}
\newcommand{\Qtil}{\ensuremath{\widetilde{Q}}}

\newcommand{\rH}{\ensuremath{r_h}}

\newcommand{\bmax}{\ensuremath{b_\mathrm{max}}}
\newcommand{\Zin}{\ensuremath{Z_\mathrm{in}}}
\newcommand{\Zout}{\ensuremath{Z_\mathrm{out}}}
\newcommand{\Yin}{\ensuremath{Y_\mathrm{in}}}
\newcommand{\Yout}{\ensuremath{Y_\mathrm{out}}}
\begin{document}
\begin{titlepage}
\begin{center}
{\hbox to\hsize{\hfill KEK-TH-828}}
{\hbox to\hsize{\hfill KIAS P02074}}
{\hbox to\hsize{\hfill TUM-HEP-429/02}}
{\hbox to\hsize{\hfill UT-ICEPP 02-07}}

\bigskip
\bigskip

\textbf{\Large
Rotating black holes at future colliders:\\
Greybody factors for brane fields
}

\bigskip
\bigskip

\textbf{Daisuke Ida}\\
\smallskip
\textit{\small
Department of Physics, Tokyo Institute of Technology, \\
Tokyo 152-8551, Japan}

\medskip

\textbf{Kin-ya Oda}\\
\smallskip
\textit{\small
Physik Dept.\ T30e, TU M\"unchen, \\
James Franck Str., D-85748 Garching, Germany}

\medskip

\textbf{Seong Chan Park}\\
\smallskip
\textit{\small
Korea Institute for Advanced Study (KIAS), \\
Seoul 130-012, Korea}

\bigskip

\texttt{d.ida@th.phys.titech.ac.jp}\\
\texttt{odakin@ph.tum.de}\\
\texttt{spark@kias.re.kr}

\bigskip
\bigskip

%
\textbf{Abstract}\\
\end{center}
\noindent \small
We study theoretical aspects of the rotating black hole
production and evaporation
in the extra dimension scenarios with TeV scale gravity,
within the mass range
in which the higher dimensional Kerr solution provides
good description.
We evaluate the production cross section of black holes
taking their angular momenta into account.
We find that it becomes larger
than the Schwarzschild radius squared,
which is conventionally utilized in literature,
and our result nicely agrees with the recent numerical study by
Yoshino and Nambu within a few \% error for higher dimensional case.
In the same approximation to obtain the above result,
we find that
the production cross section becomes {\em larger}
for the black hole with larger angular momentum.
Second, we derive the generalized Teukolsky equation
for spin 0, 1/2 and 1 brane fields
in the higher dimensional Kerr geometry
and explicitly show that it is separable in any dimensions.
For five-dimensional (Randall-Sundrum) black hole,
we obtain analytic formulae for the greybody factors
in low frequency expansion and
we present the power spectra of the Hawking
radiation as well as their angular dependence.
Phenomenological implications of our result are briefly sketched.

\end{titlepage}

\section{Introduction}
The fundamental gravitational scale can be lowered down to TeV scale
to remedy the hierarchy between the Planck and Higgs mass scales
in the large extra dimension (ADD) scenario~\cite{Arkani-Hamed:1998rs}
(see also ref.~\cite{Antoniadis:1998ig} for its stringy realization).~%
\footnote{
When the number of extra dimensions is two
(and hence their size is around mm),
rather stringent cosmological constraint $M\gtrsim 100\,\mathrm{TeV}$ is
imposed~\cite{Hall:1999mk}.
}
In warped compactification (Randall-Sundrum) scenario,
both of them are scaling together along the location in the
warped extra dimension, leading again to the TeV fundamental scale
at our visible brane~\cite{Randall:1999ee}.
When nature realizes such a TeV scale gravity scenario,
it is predicted that black hole production will dominate over the two body
scattering well above the fundamental scale,
with the geometrical cross section of the order of the 
Schwarz\-schild radius
squared (of the black hole mass equal to the center
of mass (CM) energy of the scattering)~\cite{Banks:1999gd}.
Following the observation that black holes will mainly decay into the
standard model fields on the brane rather than into
the bulk modes~\cite{Emparan:2000rs},
collider signatures of black hole production and evaporation
are studied comprehensively in ref.~\cite{Giddings:2001bu}
and independently in ref.~\cite{Dimopoulos:2001hw}.~%
\footnote{
See also ref.~\cite{Argyres:1998qn} for the study before this observation.
}
These two pioneering works
are applied in a lot of papers of the black hole phenomenology
of the ultra-high energy cosmic neutrino
signature~\cite{UHEneutrino,Anchordoqui:2002fc} and
of the collider signatures~\cite{BHcollider,Han:2002yy,Anchordoqui:2002cp}.
(In ref.~\cite{Voloshin:2001vs}, it is claimed that
the black hole production cross section would be exponentially suppressed
rather than being geometrical;
this is later answered by the semi-classical argument~\cite{Giddings:2001ih}
\footnote{
Further claim that the classical black hole formation
in the two body scattering is proved only with zero impact
parameter~\cite{Voloshin:2001fe}
is answered in refs.~\cite{Eardley:2002re,Kohlprath:2002yh,Yoshino:2002tx}.
}
and by the correspondence principle applied
to the production cross sections of black holes and
strings~\cite{Dimopoulos:2001qe}.
\footnote{
We may observe similar correspondence
in the power of the exponential suppression of
the hard scattering cross section~\cite{KangOda},
following the argument in ref.~\cite{Oda:2001uw}.
})

The black hole phenomenology opens up the fascinating possibility of the
{\em experimental investigation of the quantum gravity} in the following sense.
As is emphasized in ref.~\cite{Giddings:2001bu},
the black hole production hides all the shorter distance processes than
the Planck length scale behind the event horizon
and there emerges infrared-ultraviolet duality, i.e.,
the larger the CM energy becomes, the better the semi-classical
treatment~\cite{Hawking:1975sw} of the resultant black hole is
(since the Hawking temperature of it becomes lower).
In string theory, where its non-perturbative definition is not yet established,
this kind of situation (duality) often appears so that
one picture is valid in one limit while the other is valid in the
opposite limit (see e.g.\ ref.~\cite{Polchinski:1996nb}
for review and also refs.~\cite{Horowitz:1997nw,Dimopoulos:2001qe}).
The region of true interest is the intermediate one at which
both pictures break down and non-perturbative formulation of the
quantum gravity (or string theory) becomes relevant.
Given the status of the theoretical development,
experimental signature of quantum gravity at this intermediate region
would be observed as the discrepancy
from the semi-classical behavior in the black hole picture
valid at high energy limit.
Therefore in order to investigate the quantum gravity effect,
it is essential to predict this semi-classical behavior
as precisely as possible. This is the main motivation of our work.

After the production phase (the ``balding'' phase),
the black holes are well described by the higher dimensional Kerr
solution~\cite{Myers:1986un} if the mass of the produced black hole
($\simeq$ the CM energy of the collision) is
large enough to neglect the brane tension at the horizon and also
small enough to neglect the topology and the curvature of the extra
dimension(s)~\cite{Giddings:2001bu}.
Within the LHC energy
region,
the former condition is satisfied (or marginal)
and the latter is perfect in the ADD scenario~\cite{Giddings:2001bu,Dimopoulos:2001hw}
while the former is the same as
in the ADD scenario and the latter is satisfied
in the Randall-Sundrum scenario
(when the horizon radius is smaller than the curvature length scale
which is one or two order(s) of magnitude
larger than the Planck length scale
\footnote{
The refs.~\cite{Giddings:2000mu,Giddings:2000ay,Arkani-Hamed:2000ds}
considering mainly the application
of the AdS-CFT correspondence also support this view.
})
~\cite{Giddings:2001bu,Anchordoqui:2002fc}.
Throughout this paper, we assume that both the two conditions are satisfied.

The black hole emits most of its quanta
(and hence loses most of its mass and angular momentum)
through the Hawking radiation~\cite{Hawking:1975sw}
when the above ``large-enough'' (former) condition is satisfied
and hence a few hot quanta emitted in the final ``Planck'' phase,
which cannot be treated
semiclassically,
does not consist of the main part of the decay product~\cite{Giddings:2001bu}.
(Remember that the smaller the black hole becomes,
the hotter the Hawking radiation.)
\nocite{Page:1976df,Page:1976ki}
In most literature
the ``spin-down''
phase of the black hole evolution~\cite{Giddings:2001bu},
in which the black hole shed its
angular momentum, is simply neglected and
the Schwarz\-schild black hole is used from the start
relying on the four-dimensional result~\cite{Page:1976ki} that
the half life for spin down is a few \% of the black hole lifetime.
\footnote{
The spin-down phase accounts for about 25\% of the mass loss
in this four-dimensional case~\cite{Page:1976ki}.
}
To improve this ``Schwarz\-schild approximation'',
it is important to estimate the production cross section of the black holes
with finite angular momenta.
In ref.~\cite{Kotwal:2002wg}, the production cross section of rotating black
holes is estimated from the quantum mechanical matrix element between
the initial two-plane-waves state
and the `black hole state'.
In this paper, we take more conservative approach based on the
(classical) geometrical cross section,~\footnote{
See also ref.~\cite{Park:2001xc} for earlier heuristic
attempt to estimate the rotating black hole production.
}
in the spirit~\cite{Giddings:2001ih} that
a classical description
should be more or less valid for the black hole production
in order to avoid the Voloshin's exponential suppression mentioned above.
\footnote{
See refs.~\cite{Solodukhin:2002ui,Hsu:2002bd} for the quantum
argument which also claim that Voloshin's suppression is not applied.
}

The Hawking radiation is determined for each mode by the greybody
factor~\cite{Hawking:1975sw,Page:1976df},
i.e.\ the absorption probability (by the black hole)
of an in\-com\-ing wave of the corresponding mode.
\footnote{
It is
first calculated
for spin~0 field~\cite{Starobinsky:I}, then for spin~1 and~2
fields~\cite{Teukolsky:1972my,Starobinsky:II,Teukolsky:I,Teukolsky:II,Teukolsky:III},
and finally for spin~1/2 field~\cite{Page:1976df,Page:1976ki}
for four-dimensional Kerr black hole.
}
Unfortunately, the greybody factors
have been calculated only for brane- and bulk-scalar modes
with the Schwarz\-schild black hole at present~\cite{Kanti:2002nr}.
In current black hole phenomenology,
the Hawking radiation is
either not considered (typically in the cosmic neu\-tri\-no signature)
or considered with the greybody
factor in the geo\-met\-ric\-al op\-tics limit.
\footnote{
See e.g.\ refs.~\cite{Han:2002yy,Anchordoqui:2002cp} for
consideration of the greybody factor in the geometrical optics limit
for higher dimensional black hole.
}
%
To study the evaporation of the higher dimensional black hole
and to progress the phenomenology further,
it is prerequisite to obtain the greybody factors
of the brane fields
(which are the main decay modes of the black hole as is mentioned above).

In this paper, we obtain the brane field equations
generalizing the Teukolsky's method in
four dimensions~\cite{Teukolsky:1972my,Teukolsky:I,Teukolsky:II,Teukolsky:III}.
We show that they are separable into radial and angular parts.
For the five-dimensional Kerr black hole,
we find the analytic formulae of the greybody factors
within the low frequency expansion.

In section~2, we present the geometrical production cross section of
rotating black holes with finite angular momenta
in the approximation neglecting the balding phase.
Our result of the largest impact parameter $\bmax$
for the black hole formation turns out to be
in good agreement with the numerical result by
Yoshino and Nambu~\cite{Yoshino:2002tx}.
Within the same approximation,
we find that the (differential) cross section linearly increases
with the angular momentum for a given black hole mass ($\simeq$ CM energy).
We also estimate the production of the exploding black ring
and find that it will possibly form when there are many extra dimensions.
In section~3,
we study the Hawking radiation from the rotating black hole.
First we derive the brane field equations for
the spin 0, 1/2, and 1 brane fields from the induced metric
on the brane in the higher dimensional Kerr black hole back ground
and show that these equations are separable into radial and angular parts
for any number of extra dimensions.
Next
we find the analytic expression for the greybody factors for brane fields
for the rotating five-dimensional (Randall-Sundrum) black hole
within the low frequency expansions.
We present the power spectra as well as their angular dependence
applying these greybody factors.
In section~4, we present a summary and briefly
comment on the phenomenological
implications of our results.

\section{Production of rotating black holes}
First we briefly review the properties of the rotating $(4+n)$- dimensional black
hole~\cite{Myers:1986un}.
Since we assume that the
large enough condition (explained in the Introduction)
is satisfied, the charges of the black hole can be neglected;
they are at most a few coming from the initial two particles.
In general,
higher dimensional black hole may have $\lfloor(n+3)/2\rfloor$
angular momenta.
When the black hole is produced in the collision of two
particles on the brane, where
the initial state has only single angular momentum (directed in the brane),
it is sufficient to consider that the only single angular momentum is
non-zero~\cite{Giddings:2001bu}.
(This implicitly assumes that the balding phase can be neglected,
namely that the ``junk'' emissions are negligible and
do not change the $\lfloor(n+3)/2\rfloor$ angular momenta during this phase.)
In the Boyer-Lindquist coordinate,
the metric for the black hole with single angular momentum
takes the following form~\cite{Myers:1986un}
\begin{eqnarray}
g &=&\left(1-\frac{\mu r^{-n+1}}{\Sigma(r,\vartheta)}\right)dt^2
  -\sin^2\vartheta\left(
    r^2+a^2+a^2\sin^2\vartheta\frac{\mu r^{-n+1}}{\Sigma(r,\vartheta)}\right)d\varphi^2
    \nonumber\\
&&\mbox{}+2a\sin^2\vartheta\frac{\mu
r^{-n+1}}{\Sigma(r,\vartheta)}dtd\varphi
  -\frac{\Sigma(r,\vartheta)}{\Delta(r)}dr^2-\Sigma(r,\vartheta) d\vartheta^2\nonumber\\
&&\mbox{}-r^2\cos^2\vartheta\,d\Omega^n, \label{eq:metric_general}
\end{eqnarray}
where
\begin{eqnarray}
\Sigma(r,\vartheta)&=&r^2+a^2\cos^2\vartheta,\nonumber\\
\Delta(r)       &=&r^2+a^2-\mu r^{-n+1}.\nonumber
\end{eqnarray}
We can see that the horizon occurs when $\Delta(r)=0$, i.e.\ when $r=r_h$ with
\begin{eqnarray}
r_h &=&
\left[\frac{\mu}{1+\astar^2}\right]^{1/(n+1)}
= (1+\astar^2)^{-1/(n+1)}r_S,\label{eq:r_h}
\end{eqnarray}
where $\astar=a/r_h$ and the Schwarz\-schild radius
$r_S=\mu^{1/(n+1)}$
are introduced for later convenience.
Note that there is only single horizon when $n\geq 1$
(contrary to the four-dimensional Kerr black hole which has inner
and outer horizons)
and its radius is independent of the angular coordinates.
We can obtain the total mass
 $M$ and angular momentum $J$ of the black hole
from the metric (\ref{eq:metric_general})
\begin{eqnarray}
M = \frac{(n+2)A_{n+2}}{16\pi G}\mu,
\hspace{1cm}
J = \frac{2}{n+2}Ma,\label{eq:MandJ}
\end{eqnarray}
where $A_{n+2}$$=$$2\,\pi^{(n+3)/2}/\Gamma(\frac{n+3}{2})$ is the
area of unit sphere $S^{n+2}$ and $G$ is  the $(4+n)$-dimensional
Newton constant. Therefore we may consider $\mu$ and $a$ (or
$r_h^{-1}$ and $\astar$) as the normalized mass and angular
momentum parameters, respectively. We note that there are no upper
bound on $a$ when $n\geq 2$ nor on $\astar$ when $n\geq 1$,
contrary to the four-dimensional case where both $a$ and $\astar$
are bounded from above. In this paper, we concentrate on the brane
field equations and hence only the induced metric on the brane is
relevant, where the last term in eq.~(\ref{eq:metric_general})
vanishes and the angular variables $\vartheta$ and $\varphi$ are
redefined to take the values $0\leq\vartheta\leq\pi$ and
$0\leq\varphi<2\pi$.
The explicit form is given in eq.~(\ref{eq:brane_metric}).

\subsection{Production cross section}
\begin{figure}
\begin{center}
\epsfig{file=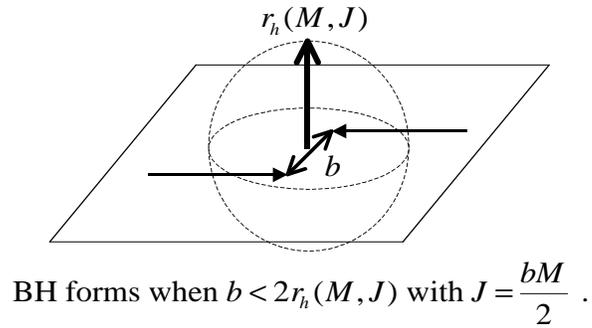,width=12cm}
\caption{
Schematic picture for the condition of the black hole formation.
} \label{fig:BHpicture}
\end{center}
\end{figure}

We estimate the production cross section of
rotating black holes within the classical picture.
Let us consider a collision of two massless particles
with finite impact parameter $b$ and CM energy $\sqrt{s}=M_i$
so that each particle has energy $M_i/2$ in the CM frame.
The initial angular momentum before collision is $J_i=bM_i/2$ (in the CM frame).
Suppose that a black hole forms
whenever the initial two particles (characterized by $M_i$ and $J_i$)
can be wrapped inside the event horizon of
the black hole with the mass $M=M_i$ and angular momentum $J=J_i$
(see Fig.~\ref{fig:BHpicture} for schematic picture),
i.e., when
\begin{eqnarray}
b &<& 2r_h(M,J)=2r_h(M_i,bM_i/2), \label{eq:our_condition}
\end{eqnarray}
where $r_h(M,J)$ is defined through eqs.~(\ref{eq:r_h}) and (\ref{eq:MandJ}).
Since the right hand side is monotonically decreasing function of $b$,
there is maximum value $\bmax$ which saturates the
inequality~(\ref{eq:our_condition})
\begin{eqnarray}
\bmax(M)=2\left[1+\left(\frac{n+2}{2}\right)^2\right]^{-{1\over n+1}}r_S(M),
\label{eq:bmax_formula}
\end{eqnarray}
where $r_S(M)$ is defined by $r_S(M)=\mu(M)^{1/(n+1)}$ and eq.~(\ref{eq:MandJ}).
When $b=\bmax$, the rotation parameter $\astar$ takes the maximal value
$(\astar)_\mathrm{max}=(n+2)/2$.

The formula (\ref{eq:bmax_formula})
fits the numerical result of $\bmax$
with full consideration of the general relativity
by Yoshino and Nambu~\cite{Yoshino:2002tx}
within the accuracy
less than 1.5\% for $n\geq 2$ and 6.5\% for $n=1$
(although it just gives the Schwarz\-schild radius $\bmax=r_S(M)$ for $n=0$
which is 24\% larger than the numerical result~\cite{Eardley:2002re}):
\begin{eqnarray}
\begin{array}{c|cccccccc}
n  & 0 & 1 & 2 & 3 & 4 & 5 & 6 & 7 \\
\hline
R_\mathrm{Numerical}~\cite{Yoshino:2002tx}
   & 0.804 & 1.04 & 1.16 & 1.23 & 1.28 & 1.32 & 1.35 & 1.37\\
R_\mathrm{Analytic}
   & 1.00  & 1.11 & 1.17 & 1.22 & 1.26 & 1.30 & 1.33 & 1.36
\end{array}, \nonumber
\end{eqnarray}
where $R$ denotes $R=\bmax/r_S(M)$.

Our result
is obtained in the 
approximation that we neglect all the effects by the junk emissions
in the balding phase
and hence that
the initial CM energy $M_i$ and angular momentum $J_i$ become directly
the resultant black hole mass $M=M_i$ and angular momentum $J=J_i$.
\footnote{
The authors of ref.~\cite{Yoshino:2002tx} have found that
the irreducible mass of the black hole
is substantially reduced when $b$ is close to $\bmax$
and have suggested that balding phase is not negligible when $b\sim\bmax$.
However, the irreducible mass
provides the lower bound on the final mass
of the black hole;
at this stage we cannot conclude how much junk energy and angular
momentum are radiated at the balding phase.
}
The coincidence of our result with the numerical study~\cite{Yoshino:2002tx}
suggests that this approximation would be actually viable
for higher dimensional black hole formation
at least unless $b$ is very close to $\bmax$.~%
\footnote{
See refs.~\cite{Cardoso:2002ay,Cardoso:2002yj,Cardoso:2002jr}
for estimation of the energy loss during the balding phase
for the head-on collision (b=0) case
obtained from gravitational radiation
emitted during the infall of a particle into a four dimensional black hole.
}

Once we neglect the balding phase, hence the junk emission,
the initial impact parameter $b$ directly leads to
the resultant angular momentum of the black hole $J=bM/2$.
Since the impact parameter $[b, b+db]$ contributes to the
cross section $2\pi bdb$,
this relation between $b$ and $J$
tells us the (differential) production cross section
of the black hole with its mass $M$ and its angular momentum in $[J,J+dJ]$
\begin{eqnarray}
d\sigma(M,J)=\left\{
   \begin{array}{cc}
   8\pi JdJ/M^2 & (J<J_\mathrm{max})\\
   0            & (J>J_\mathrm{max})
   \end{array}\right.
, \label{eq:dsigma_dJ}
\end{eqnarray}
where
\begin{eqnarray}
J_\mathrm{max}
   &=&\frac{\bmax M}{2}=j_n\,\left(\frac{M}{M_P}\right)^{n+2\over n+1}
\end{eqnarray}
with
\footnote{
See e.g.\ ref.~\cite{Giddings:2001ih} for different conventions for $M_P$.
}
\begin{eqnarray}
j_n=\left[2^n\pi^{n-3\over 2}\Gamma\left(n+3\over 2\right) \over (n+2)
   \left[1+\left(n+2\over 2\right)^2\right]\right]^{1/(n+1)},
\hspace{1cm}
M_P=\left((2\pi)^n\over 8\pi G\right)^{1/(n+2)}.
\end{eqnarray}
The numerical values for $j_n$ are summarized in Table~\ref{table:jandk}.
\begin{table}[htbp]
\center
\caption{Numerical values for $j_n$ and $k_n$ \label{table:jandk}}
\begin{eqnarray*}
   \begin{array}{|c|cccccccc|}
   \hline
   n  &0     &1    &2    &3    &4    &5    &6    &7     \\
   \hline
   j_n    &0.0398&0.256&0.531&0.815&1.09 &1.37 &1.63  &1.88  \\
   k_n    &0.0159&0.125&0.228&0.251&0.214&0.155&0.101 &0.0603\\
   k_n/j_n&0.399 &0.489&0.429&0.308&0.195&0.114&0.0619&0.0320\\
   \hline
   \end{array}
\end{eqnarray*}
\end{table}

It is observed that the differential cross section~(\ref{eq:dsigma_dJ})
linearly increases with the angular momentum.
We expect that this behavior is correct as the first approximation,
so that
the black holes tend to be produced with larger angular momenta.
At the typical LHC energy $M/M_P=5$, the value of $J_\mathrm{max}$
is $J_\mathrm{max}=2.9, 4.5, \ldots, 10, 12$\ for
$n=1, 2, \ldots, 6,7$, respectively.
This means that the semi-classical treatment of
the angular momentum becomes increasingly valid for
large $n$.

Integrating the expression~(\ref{eq:dsigma_dJ}) simply gives
\begin{eqnarray}
\sigma(M)
&=&\pi\bmax^2
   =4\left[1+\left(\frac{n+2}{2}\right)^2\right]^{-2/(n+1)}\,\pi r_S(M)^2
   \nonumber\\
&=&F\,\pi r_S(M)^2.
\end{eqnarray}
The factor $F$ is summarized as
\begin{eqnarray}
\begin{array}{c|cccccccc}
n  & 0 & 1 & 2 & 3 & 4 & 5 & 6 & 7 \\
\hline
F_\mathrm{Numerical}~\cite{Yoshino:2002tx}
   & 0.647 & 1.084 & 1.341 & 1.515 & 1.642 & 1.741 & 1.819 & 1.883\\
F_\mathrm{Analytic}
   & 1.000 & 1.231 & 1.368 & 1.486 & 1.592 & 1.690 & 1.780 & 1.863
\end{array}. \label{eq:Rsq_table} \nonumber
\end{eqnarray}
This result implies that,
apart from the four-dimensional case,
we would underestimate the production cross section of black holes
if we did not take the angular momentum into account
and that it becomes more significant for higher dimensions.
We point out that this effect has been often overlooked in the literature.


\subsection{Rotating black ring}
In four dimensions, the topology of the event horizon must be homeomorphic to two-sphere
and there is a uniqueness theorems for static or stationary black holes.
On the other hand, a higher-dimensional black hole can have various
nontrivial topology~\cite{Cai:2001su}, and the uniqueness property of stationary black holes
fails in five (and probably in higher) dimensions.
The typical example in five dimensions
has been recently given by Emparan and Reall~\cite{Emparan:2001wn}.
They have explicitly provided a solution of the five-dimensional vacuum
Einstein equation, which represents the stationary rotating black ring
(homeomorphic to $S^1\times S^2$).
In this case, the centrifugal force prevents the black ring from collapsing.
When the angular momentum is not large enough, the black ring will
collapse to the Kerr black hole due to
the gravitational attraction and some effective tension of
the ring source. In fact, this five dimensional
black ring solution has the minimum
possible value of the angular momentum given by
\begin{eqnarray}
J_\mathrm{min}=k_\mathrm{BR}\left(M\over M_P\right)^{3/2},
\end{eqnarray}
where $k_\mathrm{BR}=0.282$.
On the other hand, we have the upper bound for the angular momentum
of the black holes produced by particle collisions:
\begin{eqnarray}
J_\mathrm{max}=j_1\left(M\over M_P\right)^{3/2},
\end{eqnarray}
where $j_1=0.256$.
Since these numerical values are of the same order, 
we cannot conclude
the possibility of black ring productions at colliders.
\footnote{Even if
the black rings are produced, they might be unstable
due to the existence of $J_\mathrm{min}$ and the black string instability.
D.I.\ is indebted to Roberto Emparan for this point.
}

Here we consider the possibility of the higher dimensional black ring
which is homeomorphic to $S^1\times S^n$.
Corresponding Newtonian situation will be the system of a rotating massive circle.
They are always unstable in higher dimensions; a circle with slow rotation
collapses and one with rapid rotation explodes toward
infinitely large thin circle.
In general relativity, we have no idea as to the validity of this picture
due to the nonlinearity of the Einstein equation.
We shall discuss in the following the possibility of the black ring formation
based on Newtonian picture assuming
that the nonlinear effects of the gravity unchange the qualitative feature.
For simplicity, we just consider the gravitational attraction and the centrifugal force
of the massive circle and neglect the effect of tension.
Let $\ell$, $M$ and $J$ be the radius, the mass and the angular momentum
 of the massive circle.
Then we obtain ($3+n$)-dimensional effective theory
with the Newton constant $G/2\pi\ell$ by integrating out
along the $S^1$-direction. The Schwarzschild radius of
the point mass in the effective theory is given by
\begin{eqnarray}
r &\sim& \left[\frac{16\pi G}{(n+1)A_{n+1}}\frac{M}{2\pi l}\right]^{1/n}
= \left[\frac{8GM}{(n+1)lA_{n+1}}\right]^{1/n}.
\label{sn-radius}
\end{eqnarray}
Thus we expect the black ring with $S^1$-radius $\ell$ and $S^n$-radius $r$.
In flat space picture, $\ell>r$ should hold for black ring.
This condition gives
\begin{eqnarray}
\ell &\gtrsim& \ell_\mathrm{min}=\left[8GM\over (n+1)A_{n+1}\right]^{1/(n+1)}.
\label{donatsu}
\end{eqnarray}
On the other hand, the condition that the centrifugal force dominates
against the gravitational attraction becomes
\begin{eqnarray}
J &\gtrsim& 2^{-(n+3)/2}G^{1/2}\ell^{-(n-1)/2}M^{3/2}.
\end{eqnarray}
This combined with eq.~(\ref{donatsu}) gives the minimum value of the
angular momentum for exploding black ring:
\begin{eqnarray}
J &\gtrsim& J_\mathrm{min} =k_n\left(M\over M_P\right)^{(n+2)/(n+1)},
\end{eqnarray}
where
\begin{eqnarray}
k_n &=& 2^{-{2n^2+3n+7\over 2(n+1)}}\pi^{(n+6)(n-1)\over 4(n+1)}
\left[\Gamma\left(n+2\over 2\right)\over n+1\right]^{-{n-1\over 2(n+1)}}.
\end{eqnarray}
The numerical values for $k_n$ are presented in Table~\ref{table:jandk}.
This result shows that $J_\mathrm{min}$ for exploding black rings
is one or two order(s) of magnitude 
smaller than $J_\mathrm{max}$ for collision limit
when $n$ is large.
Therefore we expect that the exploding black rings are possibly
produced at colliders if there are many extra dimensions,
though they will suffer from the black string instability when they become
sufficiently large thin rings.
In the following of this paper, we do not follow the evolution of the
exploding black ring nor consider the radiations from it
since this is still at the heuristic stage;
we concentrate on the Hawking radiations
from the higher dimensional Kerr black hole
after the balding phase.

\section{Radiations from rotating black hole}
In this section, we study the Hawking radiation~\cite{Hawking:1975sw}
from the higher dimensional Kerr black hole~\cite{Myers:1986un}.
The Hawking radiation is thermal but not strictly black body due to
the frequency dependent greybody factor $\Gamma$, which is identical to
the absorption probability (by the hole) of the corresponding
mode~\cite{Hawking:1975sw,Page:1976df}.
The quantity $1-\Gamma$ for each mode
can be computed from the solution (to the wave equation of that mode)
having no outgoing flux at the horizon
as the ratio of the incoming and outgoing flux at infinity.

It can be shown that the higher dimensional 
black hole radiates comparable amount of energy
into one brane mode and into one bulk mode (with all the Kaluza-Klein tower
summed up)~\cite{Emparan:2000rs}.
Typically, the number of degrees of freedom is much larger 
for brane mode than for bulk mode, i.e.,
tens of the standard-model degrees of freedom are living on the brane
while there are only few degrees of freedom of the graviton
(and possibly other fields) in the bulk.
Therefore the higher dimensional black hole radiates mainly on 
the brane~\cite{Emparan:2000rs}.
For this reason, we concentrate on the greybody factors
for the brane mode in this paper.
\footnote{
We note that the bulk graviton emission may not be negligible
for highly rotating black holes since the superradiant emission is
more effective for higher spin fields~\cite{Frolov:2002as,Frolov:2002gf}.
}

\subsection{Brane field equations}
We derive the wave equations of the brane modes
using the induced four dimensional metric of
the $(4+n)$-dimensional rotating black hole~\cite{Myers:1986un}.
The wave equations
can be understood as generalization of 
the Teukolsky equation~\cite{Teukolsky:1972my,Teukolsky:I,Teukolsky:II,Teukolsky:III}
to the higher dimensional Kerr geometry.
The derivation is shown in Appendix.

We present the brane field equations for massless spin $s$ field
which are obtained from the metric (\ref{eq:metric_general})
with the standard decomposition
\begin{eqnarray}
\Phi_s=R_s(r)S(\vartheta)e^{-i\omega t+im\varphi},
\end{eqnarray}
utilizing the Newman-Penrose formalism~\cite{Newman:1962qr}
\begin{eqnarray}
&&{1\over\sin\vartheta}{d\over d\vartheta}\left(\sin\vartheta
{d S\over d\vartheta}\right)
\nonumber\\
&&{}
+\left[
(s-a\omega\cos\vartheta)^2
-
(s\cot\vartheta+m\csc\vartheta)^2
-s(s-1)+A
\right]S=0,\nonumber\\
\label{eq:angular}\\
&&\Delta^{-s}{d\over dr}\left(\Delta^{s+1}{dR\over dr}\right)+\Biggl[
{K^2\over\Delta}
+s\left(
4i\omega r
-i{[2r+(n-1)\mu r^{-n}]K\over\Delta}-n(n-1)\mu r^{-n-1}\right)
\nonumber\\
&&{}
+2ma\omega-a^2\omega^2-A
\Biggl]R=0.
\label{eq:Teukolsky}
\end{eqnarray}
where
\begin{eqnarray}
K          &=&(r^2+a^2)\omega-ma.
\end{eqnarray}

The solution of eq.~(\ref{eq:angular}) is called spin-weighted
spheroidal harmonics ${}_sS_{lm}$
(see e.g.\ ref.~\cite{Teukolsky:II,Fackerell:1977})
which reduces to the spin-weighted spherical harmonics
${}_sY_{lm}$ (see e.g.\ ref.~\cite{Goldberg:1967uu}) in the limit
$a\omega\ll 1$,
\begin{eqnarray}
{}_sS_{lm}(a\omega;\vartheta,\varphi)
   &=&{}_sY_{lm}(\vartheta,\varphi)+O(a\omega),
   \label{eq:S_by_Y}
\end{eqnarray}
where \footnote{ The so-called Condon-Shortley phase $(-1)^m$ is
inserted to reduce into the standard definition of the spherical
harmonics when $s=0$:
${}_0Y_{lm}(\vartheta,\varphi)=Y_{lm}(\vartheta,\varphi)$. }
\begin{eqnarray}
{}_sY_{lm}(\vartheta,\varphi)&=&(-1)^m e^{im\varphi}\left[
   \frac{(l+m)!(l-m)!}{(l+s)!(l-s)!}
   \frac{2l+1}{4\pi}\right]^{1/2}\left(\sin\frac{\vartheta}{2}\right)^{2l}
   \nonumber\\
&&\mbox{}
   \times\sum_j
   \left(\begin{array}{c}
      l-s \\
      j \\
   \end{array}\right)\left(\begin{array}{c}
      l+s \\
      j+s-m \\
   \end{array}\right)
   (-1)^{l-j-s} \left(\cot \frac{\vartheta}{2}\right)^{2j+s-m},\nonumber\\
   \label{eq:spherical_harmonics}
\end{eqnarray}
with the sum over $j$ being understood in the range satisfying
both $0\leq j\leq l-s$ and $0\leq j+s-m\leq l+s$.
In this limit, the eigenvalue $A$ becomes $A=A_0+O(a\omega)$
where $A_0=l(l+1)-s(s+1)$ is defined for later convenience.

We may easily check that the radial equation (\ref{eq:Teukolsky}) reduces to the
Teukolsky equation~\cite{Teukolsky:1972my,Teukolsky:I,Teukolsky:II,Teukolsky:III}
when $n=0$ (hence $\mu=2GM$).
The asymptotic solutions of eq.~(\ref{eq:Teukolsky})
at the horizon and infinity are obtained
in the same way as 
in four dimensions~\cite{Teukolsky:III}
\begin{eqnarray}
\begin{array}{cc|cc}
  \multicolumn{2}{c}{r\rightarrow\infty}&
  \multicolumn{2}{c}{r\rightarrow\rH}\\
  \mathrm{outgoing}&\mathrm{ingoing}&\mathrm{outgoing}&\mathrm{ingoing}\\
\hline
  e^{i\omega\rstar}/r^{2s+1}&e^{-i\omega\rstar}/r
 &e^{ik\rstar}              &\Delta^{-s}e^{-ik\rstar}
\end{array} \label{eq:asymptotic_solutions}
\end{eqnarray}
where
\begin{eqnarray}
k     =\omega-\frac{ma}{\rH^2+a^2},
\end{eqnarray}
and $\rstar$ is defined by $\rstar\rightarrow r$ for $r\rightarrow \infty$
and
\begin{eqnarray}
\frac{d\rstar}{dr}=\frac{r^2+a^2}{\Delta(r)}.
\end{eqnarray}

\subsection{Hawking radiation and greybody factor}
Since we have shown that the massless brane field equations are separable
into radial and angular parts,
we may write down the power spectrum of the Hawking
radiation~\cite{Hawking:1975sw}
for each massless brane mode
\begin{eqnarray}
\frac{dE_{s,l,m}}{dt\,d\omega\,d\varphi\,d\cos\vartheta}&=&
   \frac{1}{2\pi}
   \frac{{}_s\Gamma_{l,m}(r_h,a;\omega)}{e^{(\omega-m\Omega)/T}-(-1)^{2s}}
   \left|{}_sS_{lm}(a\omega;\vartheta,\varphi)\right|^2\omega,
   \label{eq:power_spectrum}
\end{eqnarray}
where
$T$ and $\Omega$ are the Hawking temperature and
the angular velocity at the horizon, respectively given by
\begin{eqnarray}
T=\frac{(n+1)+(n-1)\astar^2}{4\pi(1+\astar^2)r_h}, \hspace{1cm}
\Omega=\frac{\astar}{(1+\astar^2)r_h},
\end{eqnarray}
and ${}_s\Gamma_{l,m}(r_h,a;\omega)$ is the greybody
factor~\cite{Hawking:1975sw,Page:1976df}
which is identical to the absorption probability of the incoming wave
of the corresponding mode.
(In this paper we only consider the modes which can be treated as massless
compared with the Hawking temperature $T$
since the emissions from massive modes are Boltzmann suppressed;
Typically the standard model fields
can be treated as massless at the LHC energy range.)
Integrating eq.~(\ref{eq:power_spectrum})
by the angular variables,
we obtain
\begin{eqnarray}
\frac{dE_{s,l,m}}{dt\,d\omega}&=&
\frac{1}{2\pi}\frac{{}_s\Gamma_{lm}}{e^{(\omega-m\Omega)/T}-(-1)^{2s}}
\omega.
\label{eq:int_power_spectrum}
\end{eqnarray}
In the limit $a\omega\ll 1$ we can also write down the angular dependent power
spectrum utilizing eq.~(\ref{eq:S_by_Y})
\begin{eqnarray}
\frac{dE}{dt\,d\cos\vartheta\,d\omega}&=&
\frac{1}{2\pi}\frac{{}_s\Gamma_{lm}}{e^{(\omega-m\Omega)/T}-(-1)^{2s}}\omega
\left[\int_0^{2\pi}d\varphi|{}_sY_{lm}(\vartheta,\varphi)|^2\right]
, \label{eq:ang_dep}
\end{eqnarray}
where the integral in the square brackets can be done with
eq.~(\ref{eq:spherical_harmonics}); we summarize the
results for the leading modes in the following table.
%
\begin{eqnarray*}
\begin{array}{|ccc|c|}
\hline
s      & l & m  & \int_0^{2\pi}d\varphi|{}_sY_{lm}(\vartheta,\varphi)|^2 \\
\hline
0      & 0   &  0   & 1/2 \\
0      & 1   &  1   & (3/4)\sin^2\vartheta \\
0      & 1   &  0   & (3/2)\cos^2\vartheta \\
0      & 1   & -1   & (3/4)\sin^2\vartheta \\
\hline
1/2    & 1/2 &  1/2 & \sin^2(\vartheta/2) \\
1/2    & 1/2 & -1/2 & \cos^2(\vartheta/2) \\
\hline
1      & 1   &  1   & (3/8)(1-\cos\vartheta)^2 \\
1      & 1   &  0   & (3/4)\sin^2\vartheta \\
1      & 1   & -1   & (3/8)(1+\cos\vartheta)^2 \\
\hline
\end{array}
\end{eqnarray*}

Approximately, the time dependence of $M$ and $J$ can be determined by
\begin{eqnarray}
-\frac{d}{dt}\left(\begin{array}{c}M\\J\end{array}\right)
&=&\frac{1}{2\pi}\sum_{s,l,m} g_s\int_0^\infty d\omega
   \frac{{}_s\Gamma_{l,m}(r_h,a;\omega)}{e^{(\omega-m\Omega)/T}-(-1)^{2s}}
   \left(\begin{array}{c}\omega\\m\end{array}\right),
   \label{eq:time_dependence}
\end{eqnarray}
where $g_s$ is the number of `massless' degrees of freedom
at temperature $T$, namely the number of degrees of freedom
whose masses are smaller than $T$,
with spin $s$.
(Typically
$g_0=4$, $g_\frac{1}{2}=90$ and $g_1=24$ when $T>m_t,m_H$ and
$g_0=0$, $g_\frac{1}{2}=78$ and $g_1=18$ when $m_b<T<m_W$
for the standard model fields.)
Therefore, once we obtain the greybody factors,
we completely determine the Hawking radiation and
the subsequent evolution of the black hole up to the Planck phase,
at which the semi-classical description by the Hawking radiation
breaks down and a few quanta radiated is not predictable.

In the high frequency limit, the absorption cross section for each mode
$\sigma=(\pi/\omega^2)\Gamma$
is supposed to reach the geometrical optics
limit (see e.g.\ refs.~\cite{Han:2002yy,Anchordoqui:2002cp})
\begin{eqnarray}
\sigma_{g.o.}=  \pi \left(\frac{n+3}{2}\right)^{2/(n+1)}
\frac{n+3}{n+1}r_H^2.
\label{eq:golimit}
\end{eqnarray}
In all the phenomenological literature this limit have been applied
when one calculate the Hawking radiation.
(In the refs.~\cite{Han:2002yy,Anchordoqui:2002cp}
phenomenological weighting factors 2/3 and 1/4
are multiplied to eq.(\ref{eq:golimit})
for $s=1/2$ and $s=1$ fields, respectively, based on the result
in four dimensions~\cite{Page:1976df}.)

\subsection{Greybody factors for Randall-Sundrum black hole}
To obtain the greybody factors from eqs.~(\ref{eq:angular})
and (\ref{eq:Teukolsky}) in general dimensions, we need
the numerical calculation,
which is beyond the scope of this paper and will be shown in ref.~\cite{our_future_work}.
In this paper we present analytic expression of the greybody factors
of brane fields for $n=1$ Randall-Sundrum black hole
within the low frequency expansion.
\footnote{
See ref.~\cite{Frolov:2002xf} for the study of bulk scalar emission
of five dimensional black hole.
}
Here we outline our procedure:
First we obtain the ``near horizon'' and ``far field'' solutions in
the corresponding limits;
Then we match these two solutions
at the ``overwrapping region'' in which both limits are consistently satisfied;
Finally we impose the ``purely ingoing'' boundary condition at
the near horizon side and then read the coefficients of
``outgoing'' and ``ingoing'' modes at the far field side;
The ratio of these two coefficients
can be translated into the absorption probability
of the mode, which is nothing but the greybody factor itself.

First for convenience, we define dimensionless quantities
\begin{eqnarray}
\xi      = \frac{r-\rH}{\rH}, \hspace{1cm}
\omtil   = \rH\omega, \hspace{1cm}
\Qtil    = \frac{\omega-m \Omega}{2\pi T}
             =(1+\astar^2)\omtil-m\astar.
\end{eqnarray}
(Note that in the Schwarz\-schild limit $\astar\rightarrow 0$,
$\Qtil$ becomes $\Qtil\rightarrow \omtil$.)
Then the radial equation~(\ref{eq:Teukolsky}) becomes
\begin{eqnarray}
\xi^2(\xi+2)^2R_{,\xi\xi}+2(s+1)\xi(\xi+1)(\xi+2)R_{,\xi}
    +\widetilde VR = 0,\label{eq:Teukolsky_xi}
\end{eqnarray}
where
\begin{eqnarray}
\widetilde V
&=&\left[\omtil\xi(\xi+2)+\Qtil\right]^2
   +2is\omtil\xi(\xi+1)(\xi+2)-2is\Qtil(\xi+1)\nonumber\\
&&\mbox{}
   -[A_0+O(\astar\omtil)]\xi(\xi+2).\label{eq:potential}
\end{eqnarray}

In the near horizon
limit $\omtil\xi\ll 1$, the potential (\ref{eq:potential}) becomes
\begin{eqnarray}
\widetilde V = \Qtil^2-2is(\xi+1)\Qtil-A_0\xi(\xi+2)+O(\omtil\xi),
   \label{eq:NH_potential}
\end{eqnarray}
and the solution of eq.~(\ref{eq:Teukolsky_xi}) with
the potential~(\ref{eq:NH_potential})
is obtained with the hypergeometric function
\begin{eqnarray}
\lefteqn{R_\mathrm{NH}=}\nonumber\\
&&C_1\left(\frac{\xi}{2}\right)^{-s-\frac{i\Qtil}{2}}
      \left(1+\frac{\xi}{2}\right)^{-s+\frac{i\Qtil}{2}}
   {}_2F_1(-l-s,l-s+1,1-s-i\Qtil;-\frac{\xi}{2})\nonumber\\
&&\mbox{}
  +C_2\left(\frac{\xi}{2}\right)^{\frac{i\Qtil}{2}}
      \left(1+\frac{\xi}{2}\right)^{-s+\frac{i\Qtil}{2}}
   {}_2F_1(-l+i\Qtil,l+1+i\Qtil,1+s+i\Qtil;-\frac{\xi}{2}).\nonumber\\
\end{eqnarray}
To impose the ingoing boundary condition at
the horizon (\ref{eq:asymptotic_solutions}), i.e.\
\begin{eqnarray}
R\sim\xi^{-s}e^{-ik\rstar},
\hspace{1cm}
k\frac{d\rstar}{d\xi}\sim \frac{\Qtil}{2\xi},
\end{eqnarray}
we put $C_2=0$ and normalize $C_1=1$ without loss of generality
and then we obtain
\begin{eqnarray}
\lefteqn{R_\mathrm{NH}=}\nonumber\\
&&
   \left(\frac{\xi}{2}\right)^{-s-\frac{i\Qtil}{2}}
   \left(1+\frac{\xi}{2}\right)^{-s+\frac{i\Qtil}{2}}
   {}_2F_1(-l-s,l-s+1,1-s-i\Qtil;-\frac{\xi}{2}).\nonumber\\
   \label{eq:NHsoltion}
\end{eqnarray}

In the far field limit $\xi\gg 1+|\Qtil|$, the
eq.(\ref{eq:Teukolsky_xi}) becomes
\begin{eqnarray}
\lefteqn{0=R_{,\xi\xi}+\frac{2(s+1)}{\xi}R_{,\xi}}\nonumber\\
&&
+\left[\omtil^2+\frac{2i\omtil}{\xi}(s-2i\omtil)-\frac{1}{\xi^2}
\left[A_0+O(\omtil)\right]+O(\xi^{-3})\right]R,
\end{eqnarray}
and the solution is obtained by the Kummer's confluent hypergeometric
function
\begin{eqnarray}
R_\mathrm{FF}
&=&B_1e^{-i\omtil\xi}
   \left(\frac{\xi}{2}\right)^{l-s}
   {}_1F_1(l-s+1,2l+2;2i\omtil\xi)
   \nonumber\\
&&\mbox{}
  +B_2e^{-i\omtil\xi}\left(\frac{\xi}{2}\right)^{-l-s-1}{}_1F_1(-l-s,-2l;2i\omtil\xi),\label{eq:FFsoltion}
\end{eqnarray}
where singularity from $2l$ being integer is regularized by the higher
order terms in $\omtil$.

Matching the NH and FF solutions (\ref{eq:NHsoltion}) and (\ref{eq:FFsoltion})
in the overlapping region $1+|\Qtil|\ll \xi \ll 1/\omtil$, we obtain
\begin{eqnarray}
B_1=\frac{\Gamma(2l+1)\Gamma(1-s-i\Qtil)}{\Gamma(l-s+1)\Gamma(l+1-i\Qtil)},
\hspace{0.5cm}
B_2=\frac{\Gamma(-2l-1)\Gamma(1-s-i\Qtil)}{\Gamma(-l-s)\Gamma(-l-i\Qtil)}.
\end{eqnarray}
Then we extend the obtained FF solution toward the region $\xi\gg 1/\omtil$
\begin{eqnarray}
R_{\infty}=Y_{\rm in}e^{-i\omtil\xi}\left(\frac{\xi}{2}\right)^{-1}
          +Y_{\rm out}e^{i\omtil\xi}\left(\frac{\xi}{2}\right)^{-2s-1},
\end{eqnarray}
where
\begin{eqnarray}
Y_{\rm in}
&=&\frac{\Gamma(2l+1)\Gamma(2l+2)}{\Gamma(l-s+1)\Gamma(l+s+1)}
   \frac{\Gamma(1-s-i\Qtil)}{\Gamma(l+1-i\Qtil)}
   (-4i\omtil)^{-l+s-1}\nonumber \\
&&\mbox{}+
   \frac{\Gamma(-2l)\Gamma(-2l-1)}{\Gamma(-l-s)\Gamma(-l+s)}
   \frac{\Gamma(1-s-i\Qtil)}{\Gamma(-l-i\Qtil)}
   (-4i\omtil)^{l+s},\nonumber \\
Y_{\rm out}
&=&\frac{\Gamma(2l+1)\Gamma(2l+2)}{[\Gamma(l-s+1)]^2}
   \frac{\Gamma(1-s-i\Qtil)}{\Gamma(l+1-i\Qtil)}
   (4i\omtil)^{-l-s-1}\nonumber \\
&&\mbox{}+
   \frac{\Gamma(-2l)\Gamma(-2l-1)}{[\Gamma(-l-s)]^2}
   \frac{\Gamma(1-s-i\Qtil)}{\Gamma(-l-i\Qtil)}
   (4i\omtil)^{l-s}.
\label{eq:coeff_xi}
\end{eqnarray}

Let us define $R_{-s}$ as the solution of the equation obtained by the flip
of the sign of $s$, i.e., $s\rightarrow -s$ from eq.~(\ref{eq:Teukolsky}).
When $\Delta_{,rr}=2$ as in $n=1$ (or as in the limit $r\gg r_H$ in $n\geq 2$),
we may obtain the conserved current
in the same way as in the four-dimensional case
\begin{eqnarray}
\mathcal{J}=
\Delta\left(R_{-s}\partial_rR_s^*-R_s^*\partial_r R_{-s}\right)
   +s\Delta_{,r}R_{-s}R_s^*,
\end{eqnarray}
which satisfies $\partial_r\mathcal{J}=0$.
In the limit $r\gg r_H$,
\begin{eqnarray}
R_s      &\sim& \Yin e^{-i\omega r}r^{-1}
               +\Yout e^{i\omega r}r^{-2s-1},\nonumber\\
R_{-s}   &\sim& \Zin e^{-i\omega r}r^{-1}
               +\Zout e^{i\omega r}r^{2s-1},
\end{eqnarray}
where $\Zin=\Yin|_{s\rightarrow -s}$ and $\Zout=\Yout|_{s\rightarrow -s}$,
and $\mathcal{J}$ becomes
\begin{eqnarray}
\mathcal{J}&\sim& 2i\omega\left(\Zin\Yin^*-\Zout\Yout^*\right).
\end{eqnarray}
Therefore, we may calculate the greybody factor $\Gamma$
(=the absorption probability) in the same way as
the Page's trick~\cite{Page:1976df}
\begin{eqnarray}
\Gamma
=1-\left|\frac{Y_{\rm out}
  Z_{\rm out}}{Y_{\rm in}Z_{\rm in}}\right|
=1-\left|\frac{1-C}{1+C}\right|^2, \label{eq:Page_trick}
\end{eqnarray}
where
\begin{eqnarray}
C=\frac{(4i\omtil)^{2l+1}}{4}\left(\frac{(l+s)!(l-s)!}{(2l)!(2l+1)!}\right)^2
   \left(-i\Qtil-l\right)_{2l+1},
\end{eqnarray}
with $(\alpha)_n=\prod_{n'=1}^n(\alpha+n'-1)$
being the Pochhammer's symbol.

For concreteness, we write down
the explicit expansion of eq.~(\ref{eq:Page_trick}) up to $O(\omtil^6)$ terms
\begin{eqnarray}
{}_0\Gamma_{0,0}&=&4\omtil^2-8\omtil^4+O(\omtil^6),\nonumber\\
{}_0\Gamma_{1,m}&=&\frac{4\Qtil\omtil^3}{9}\left(1+\Qtil^2\right)
   +O(\omtil^6),\nonumber\\
{}_0\Gamma_{2,m}&=&\frac{16\Qtil\omtil^5}{2025}\left(
               1+\frac{5\Qtil^2}{4}+\frac{\Qtil^4}{4}\right)
               +O(\omtil^{10}),\nonumber
\end{eqnarray}
\begin{eqnarray}
{}_{\frac{1}{2}}\Gamma_{\frac{1}{2},m}
   &=&\omtil^2\left(1+4\Qtil^2\right)
   -\frac{\omtil^4}{2}\left(1+4\Qtil^2\right)^2
   +O(\omtil^6),\nonumber\\
{}_{\frac{1}{2}}\Gamma_{\frac{3}{2},m}
   &=&\frac{\omtil^4}{36}\left(
      1+\frac{40\Qtil^2}{9}+\frac{16\Qtil^4}{9}\right)+O(\omtil^8),
      \nonumber
\end{eqnarray}
\begin{eqnarray}
{}_1\Gamma_{1,m}
   &=&\frac{16\Qtil\omtil^3}{9}\left(1+\Qtil^2\right)+O(\omtil^6),\nonumber\\
{}_1\Gamma_{2,m}
   &=&\frac{4\Qtil\omtil^5}{225}\left(
               1+\frac{5\Qtil^2}{4}+\frac{\Qtil^4}{4}\right)
               +O(\omtil^{10}). \label{eq:Gamma_explicit}
\end{eqnarray}
Note that subleading terms in $\omtil$ are already neglected when
we obtain eq.~(\ref{eq:Page_trick}) and the terms from these contributions
are not written nor included in eqs.~(\ref{eq:Page_trick}) and (\ref{eq:Gamma_explicit}).
We also note that the so-called s-wave dominance is maximally violated
for spinor and vector fields since there are no $l=0$ modes for them.

\subsection{Radiations from Randall-Sundrum black hole}
The greybody factor (\ref{eq:Page_trick}) is obtained from low-frequency
expansions.
In four~dimensions, it is known that the greybody factors
in the low-frequency expansion provide smaller value
for the right hand side of eq.~(\ref{eq:time_dependence})
than the one from full numerical calculation~\cite{Page:1976ki}.
Therefore in this paper we do not try to show
the time evolution of the black hole nor
the time-integrated result.

\begin{figure}
\begin{center}
\epsfig{file=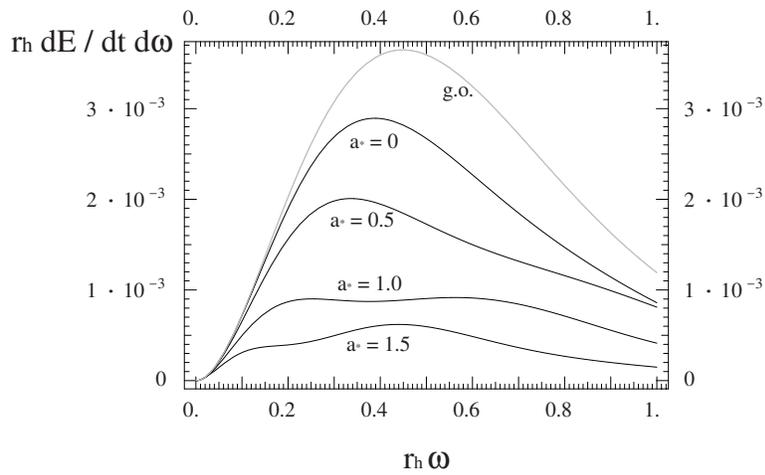,width=10cm}
\caption{
Scalar ($s=0$) power spectrum $r_hdE/dtd\omega$ vs.\ $\omtil=r_h\omega$
in lenear-linear plot.
The gray line corresponds the geometrical optics limit.
The black lines are our results for $a_*=0$, $0.5$,
$1.0$ and $1.5$ from above to below.
Note that our approximation is valid for $\omtil<\min(1,\astar^{-1})$.
} \label{fig:lins0}
\end{center}
\end{figure}

\begin{figure}
\begin{center}
\epsfig{file=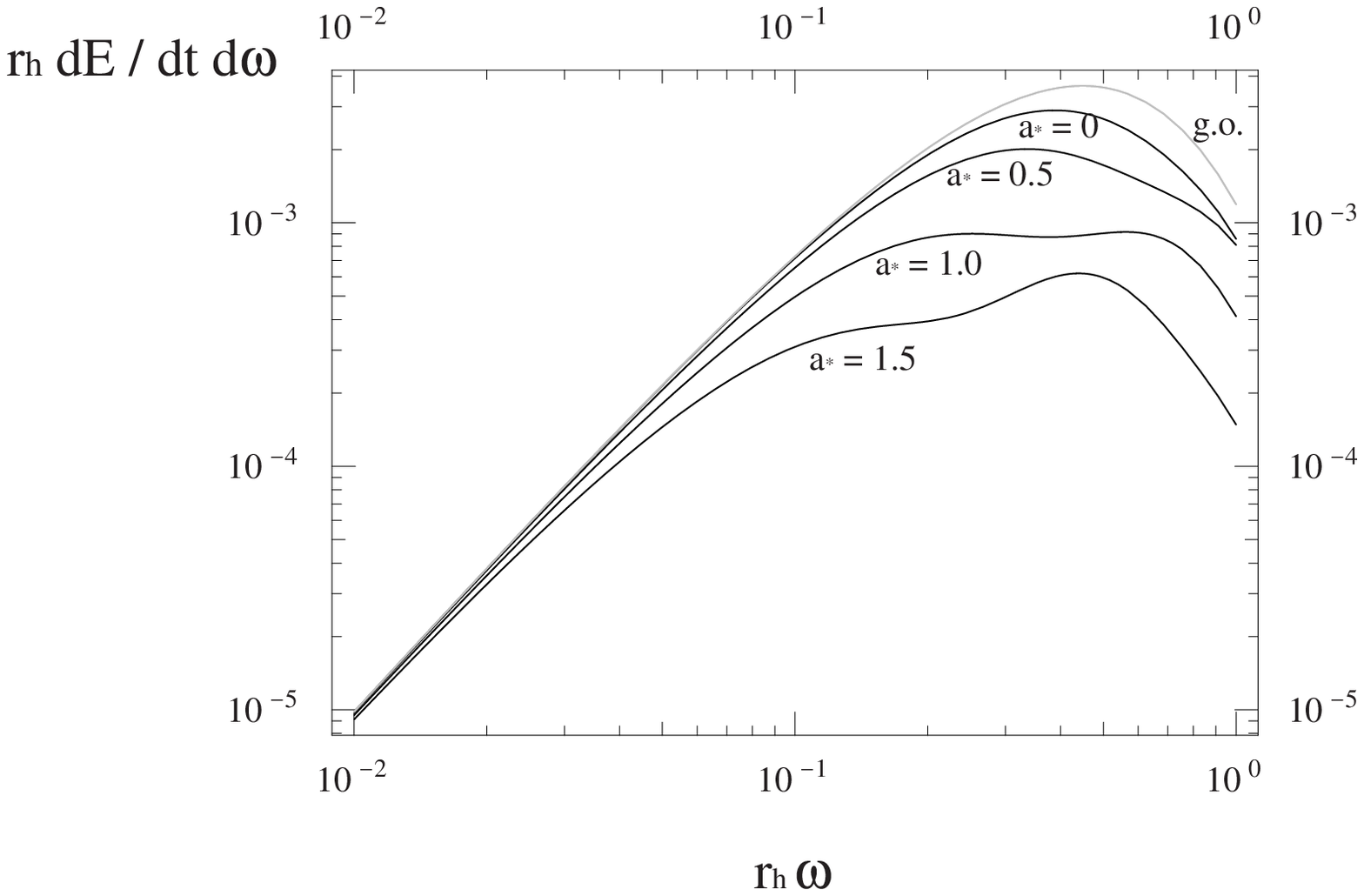,width=10cm}
\caption{
Scalar ($s=0$) power spectrum $r_hdE/dtd\omega$ vs.\ $r_h\omega$
in log-log plot.
See the caption of Fig.~\ref{fig:lins0} for explantion.
} \label{fig:logs0}
\end{center}
\end{figure}

\begin{figure}
\begin{center}
\epsfig{file=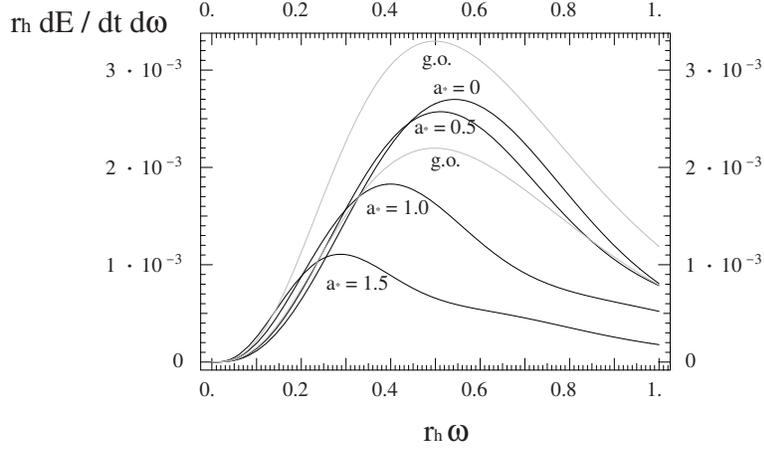,width=10cm}
\caption{
Spinor ($s=1/2$) power spectrum $r_hdE/dtd\omega$ vs.\ $\omtil=r_h\omega$
in linear-linear plot.
Two gray lines below and above
correspond to the geometrical optics limit with and without
the phenomenological weighting factor
2/3, respectively.
The black lines are our results for $a_*=0$, $0.5$,
$1.0$ and $1.5$, respectively from right to left at the peak location.
Note that our approximation is valid for $\omtil<\min(1,\astar^{-1})$.
} \label{fig:lins12}
\end{center}
\end{figure}

\begin{figure}
\begin{center}
\epsfig{file=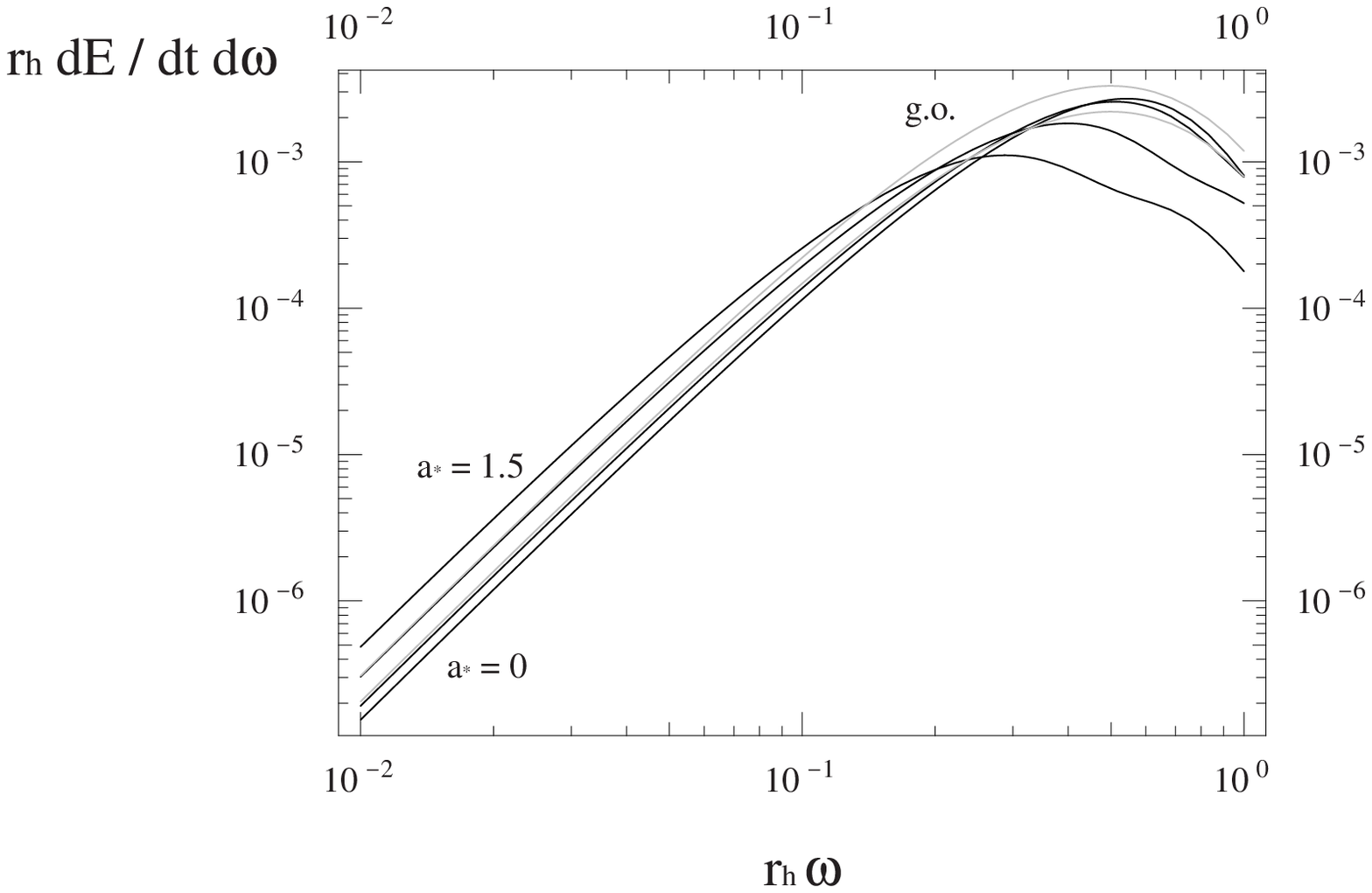,width=10cm}
\caption{
Spinor ($s=1/2$) power spectrum $r_hdE/dtd\omega$ vs.\ $r_h\omega$
in log-log plot.
See the caption of Fig.~\ref{fig:lins12} for explantion.
} \label{fig:logs12}
\end{center}
\end{figure}

\begin{figure}
\begin{center}
\epsfig{file=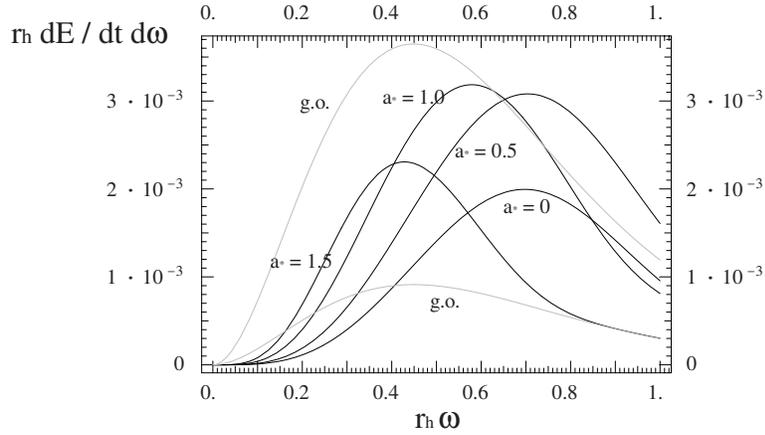,width=10cm}
\caption{
Vector ($s=1$) power spectrum $r_hdE/dtd\omega$ vs.\ $\omtil=r_h\omega$
in linear-linear plot.
Two gray lines below and above
correspond to the geometrical optics limit with and without
the phenomenological weighting factor
1/4, respectively.
The black lines are our results for $a_*=0$, $0.5$,
$1.0$ and $1.5$, respectively from below to above at the left of the peaks.
Note that our approximation is valid for $\omtil<\min(1,\astar^{-1})$.
} \label{fig:lins1}
\end{center}
\end{figure}

\begin{figure}
\begin{center}
\epsfig{file=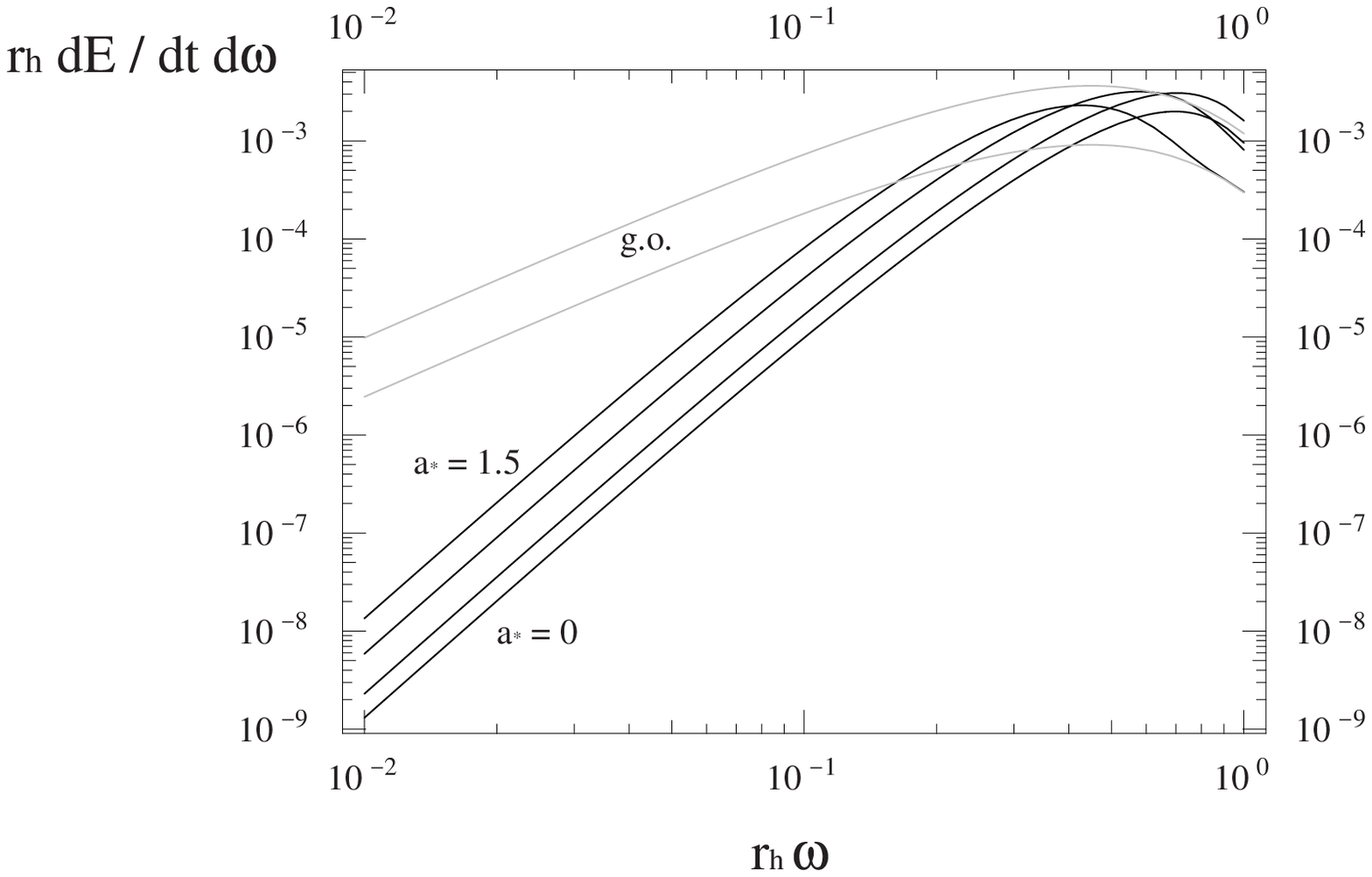,width=10cm}
\caption{
Vector ($s=1$) power spectrum $r_hdE/dtd\omega$ vs.\ $r_h\omega$
in log-log plot.
See the caption of Fig.~\ref{fig:lins1} for explantion.
} \label{fig:logs1}
\end{center}
\end{figure}

In Figs.~\ref{fig:lins0}--\ref{fig:logs1}, we show
the power spectrum~(\ref{eq:int_power_spectrum}) for spin 0,
1/2 and 1 fields.
The black lines are our results for $a_*=0$, $0.5$,
$1.0$ and $1.5$ utilizing the expression~(\ref{eq:Page_trick})
with up to $l\leq 7$ modes,
respectively from below to above at the left of the peak
(and from above to below at the right of the peak).
Note that our approximation is valid for the region
satisfying both of $a_*\omtil<1$ and $\omtil<1$,
typically at the left of the peak.
Two gray lines below and above are the corresponding power spectra
in the geometrical optics limit (\ref{eq:golimit})
with and without phenomenological weighting
factor, respectively (2/3 for spinors or 1/4 for
vectors)~\cite{Han:2002yy,Anchordoqui:2002cp}.

\begin{figure}
\begin{center}
\epsfig{file=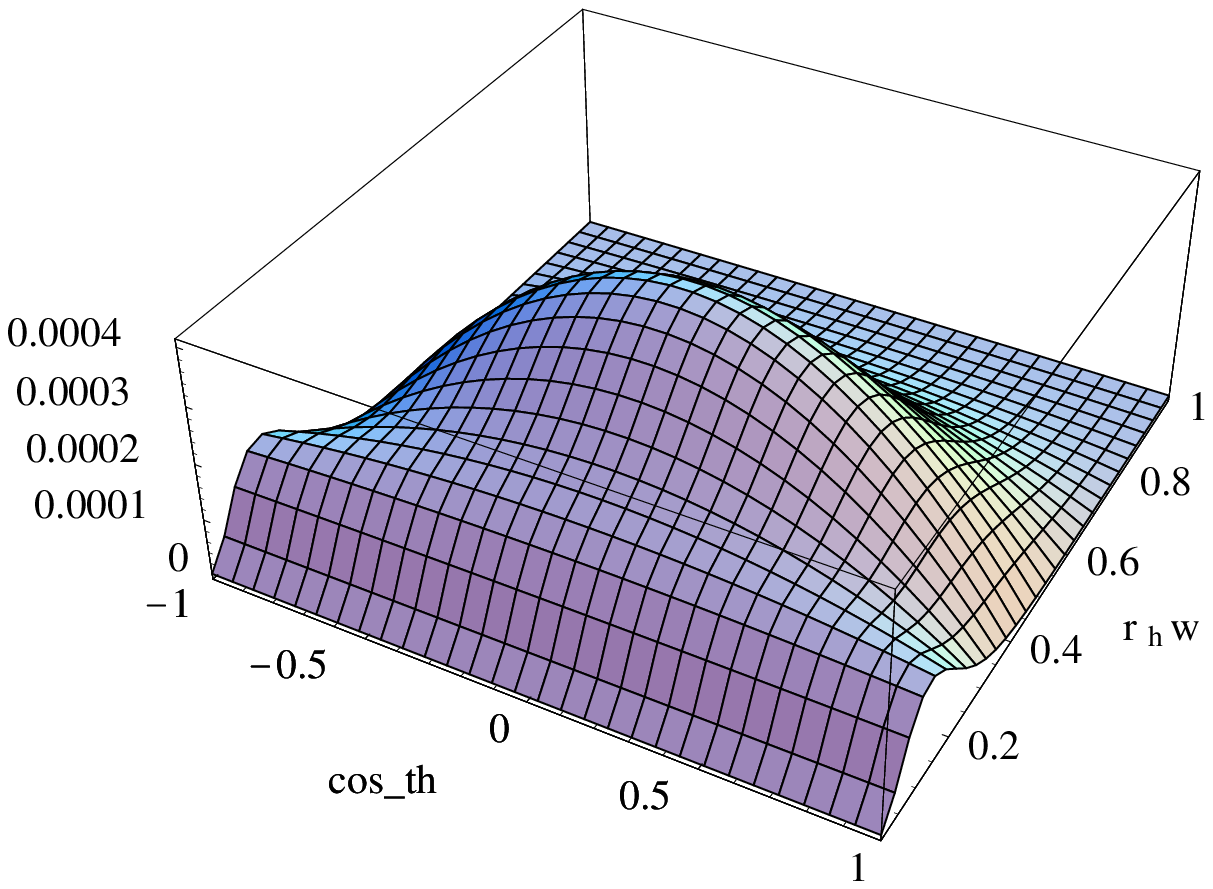,width=10cm} \caption{ Scalar ($s=0$) power
spectrum $r_hdE/dtd\omega d\cos\vartheta$ vs.\ $r_h\omega$ and
$\cos\vartheta$ for $\astar=1.5$. } \label{fig:ang0}
\end{center}
\end{figure}

\begin{figure}
\begin{center}
\epsfig{file=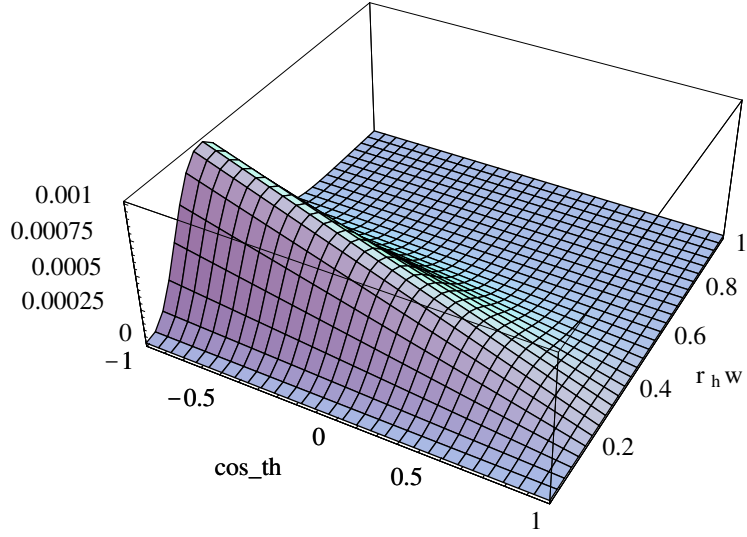,width=10cm} \caption{ Spinor ($s=1/2$)
power spectrum $r_hdE/dtd\omega d\cos\vartheta$ vs.\ $r_h\omega$
and $\cos\vartheta$ for $\astar=1.5$. } \label{fig:ang12}
\end{center}
\end{figure}

\begin{figure}
\begin{center}
\epsfig{file=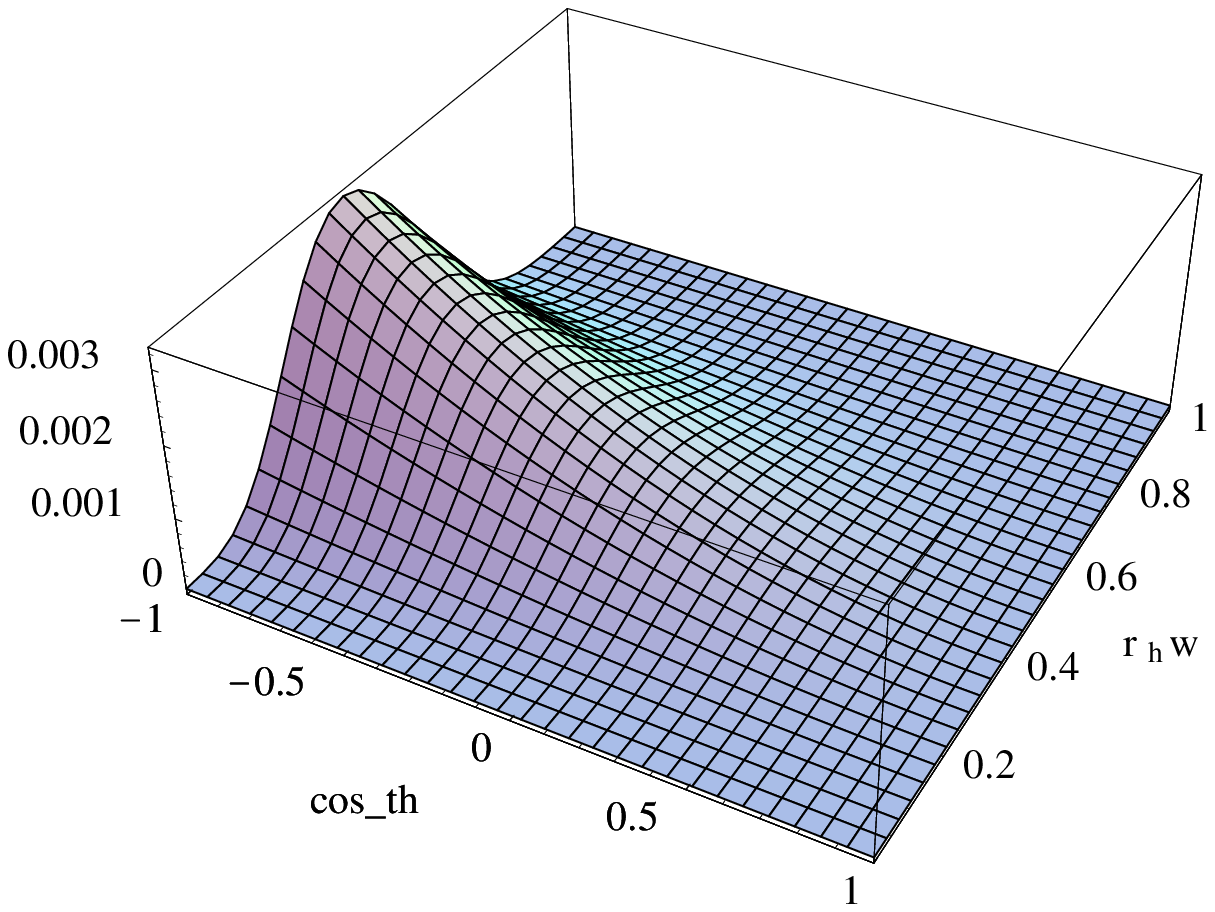,width=10cm} \caption{ Vector ($s=1$) power
spectrum $r_hdE/dtd\omega d\cos\vartheta$ vs.\ $r_h\omega$ and
$\cos\vartheta$ for $\astar=1.5$. } \label{fig:ang1}
\end{center}
\end{figure}

In Figs.~\ref{fig:ang0}--\ref{fig:ang1}, we present the angular
dependent power spectrum~(\ref{eq:ang_dep}) for spin 0, 1/2 and 1
fields when $\astar=(\astar)_\mathrm{max}=1.5$. The modes are
taken up to $l\leq 1$. We observe that there is large angular
dependence for spinors and vectors. Note that $\vartheta=0$
corresponds the direction of the angular momentum of the black
hole which is perpendicular to the beam axis. The angular
dependence shown in Figs.~\ref{fig:ang0}--\ref{fig:ang1} vanishes
when we take the limit $\astar\rightarrow 0$.

\section{Summary}
We have studied theoretical aspects of the rotating black hole production
and evaporation.

For production, we present an estimation of the geometrical cross
section up to unknown mass and angular momentum loss in the balding phase.
Our result of the maximum impact parameter $\bmax$ is in good agreement with
the numerical result by Yoshino and Nambu when the number of extra dimensions
is $n\geq 1$ (i.e.\ within 6.5\% when $n=1$ and 1.5\% when $n\geq 2$),
though ours predicts same as the
naive value in the Schwarz\-schild approximation
$\bmax=r_S(M)$ when $n=0$ which is 24\% larger than the numerical result.
(Here we note that our refinement
from the Schwarz\-schild approximation 
results in the enlargement of the production
cross section, contrary to the previous claim in the literature.)
Relying on this agreement, 
we obtain the
(differential) cross section for a given mass and (an interval of) an angular
momentum, which increases linearly with the angular momentum up to the
cut-off value $J_\mathrm{max}=\bmax M/2$.
This result shows that black holes tend to be produced with large angular
momenta.
We also studied the possibility of the black ring formation and
find that it would possibly form when there are many extra dimensions.

For evaporation, we first calculate the brane field equations for general
spin and for an arbitrary number of extra dimensions.
We show that the equations are separable into radial and angular parts
as the four-dimensional Teukolsky equations.
From these equations, we obtain the greybody factors
for brane fields with general spin 
for the five-dimensions ($n=1$) Kerr black hole
within the low-frequency expansion.
We present the resultant power spectrum
which is substantially different
from the one with geometrical approximation utilized in the literature.

We address several phenomenological implications of our results.
The production cross section of the black holes becomes larger
than the one calculated from the Schwarz\-schild radius.
The more precise determination of the
radiation power is now available.
We have shown that the black holes are produced with large
angular momenta and that the resultant radiations will have
strong angular dependence for  $s=1/2$ and $s=1$ modes
which points perpendicular to the beam axis
while very small angular dependence is expected for scalar mode.
When we average over opposite helicity states,
the up-down asymmetry with respect to the angular momentum axis shown 
in Fig.~\ref{fig:ang12}
disappears (though there still remains angular dependence itself)
\cite{Vilenkin:1978is,Leahy:1979xi};
We expect similar result for vector fields (which correspond to Fig.~\ref{fig:ang1}).
More quantitative estimation will need the greybody factors for
arbitrary frequency calculated numerically.

\bigskip
\bigskip

\begin{center}
\textbf{Acknowledgments}
\end{center}

\noindent 
We are grateful to Bob McElrath and Graham Kribs for
helpful communications, and to Gungwon Kang and Manuel Drees for
useful comments. 
D.I.\ would like to thank R. Emparan for
discussions. 
D.I.\ was supported by JSPS Research, and this
research was supported in part by the Grant-in-Aid for Scientific
Research Fund (No. 6499). 
K.O.\ thanks to the members of KIAS theory group for
their hospitality during the stay in April 2002 and is grateful to
the International Center for Elementary Particle Physics for
financial support during the main period of this work. The work of
K.O.\ is partly supported by the SFB375 of the Deutsche
Forschungs\-gemein\-schaft.

\appendix
\bigskip
\bigskip

\begin{center}
\textbf{Appendix}
\end{center}

\section{Separability of brane fields}
The various field equations in the four-dimensional Kerr background are known
to be separable. This results from the special feature of the
four-dimensional Kerr metric, that is, the vacuum metric which has
a pair of degenerate principal null directions (Petrov type D).
The four-dimensional metric considered in this paper is the
induced metric of the totally geodesic probe brane in the
higher-dimensional Kerr field. Though this brane metric turns out
to be of Petrov type D, it is not the vacuum metric itself.
Nevertheless, it happens that the massless fields on the brane are
separable, as shown bellow.

The induced metric on the three-brane in the $(4+n)$-dimensional Kerr metric (with a single nonzero angular momentum)
is given in terms of the Boyer-Lindquist coordinate system by
\begin{eqnarray}
g&=&\left(1-{\mu\over\Sigma}r^{1-n}\right)dt^2+{2a\mu\over\Sigma}r^{1-n}\sin^2\vartheta
dtd\varphi
-\sin^2\vartheta\left(r^2+a^2+{\mu a^2\sin^2\vartheta\over\Sigma}r^{1-n}\right)d\varphi^2
\nonumber\\
&&{}-{\Sigma\over\Delta}dr^2-{\Sigma}d\vartheta^2,
\label{eq:brane_metric}
\end{eqnarray}
where
\begin{eqnarray}
\Sigma=r^2+a^2\cos^2\vartheta,
\hspace{1cm}
\Delta=r^2+a^2-\mu r^{1-n},
\end{eqnarray}
and the parameters $\mu$ and $a$ are equivalent to the total mass $M$ and the angular momentum $J$
\begin{eqnarray}
M={(n+2)A_{n+2}\mu\over 16\pi G},
\hspace{1cm}
J={A_{n+2}\mu a\over 8\pi G}.
\end{eqnarray}
where $A_{n+2}=2\pi^{(n+3)/2}/\Gamma((n+3)/2)$ is the area of a unit $(n+2)$-sphere.

The direct calculation shows that the massless scalar field
equation separates on this background geometry. If we set
$\varphi=R(r)S(\vartheta)e^{-i\omega t+i m\varphi}$, then
$\nabla^2\varphi=0$ becomes
\begin{eqnarray}
&&{1\over\sin\vartheta}{d\over d\vartheta}\left(\sin\vartheta
{d S\over d\vartheta}\right)
+\left(
a^2\omega^2\cos^2\vartheta
-
m^2\csc^2\vartheta
+A
\right)S=0,\\
&&{d\over dr}\left(\Delta{dR\over dr}\right)+\Biggl[
{K^2\over\Delta}
+
4i\omega r
-i{[2r+(n-1)\mu r^{-n}]K\over\Delta}
\nonumber\\
&&{}
+2ma\omega-a^2\omega^2-A
\Biggl]R=0,
\end{eqnarray}
where $K=(r^2+a^2)\omega-am$.
We note that the Hamilton-Jacobi and massive scalar field equations
are also separable though we do not shown them here;
a test particle on the brane has an additional conserved quantity
(Carter constant) besides the energy and the angular momentum.

To show the separability of higher spinor field equations,
we work on the Newman-Penrose formalism~\cite{Newman:1962qr}.
\footnote{
See
e.g.\ ref.~\cite{Penrose:1985jw} for review of
the Newman-Penrose formalism and spinor calculations.
We follow the conventions of this reference.
}
We set the null tetrad as follows:
\begin{eqnarray}
n_\mu&=&\delta_\mu^t-a\sin^2\vartheta \delta_\mu^\varphi
-{\Sigma\over\Delta}\delta_\mu^r, \nonumber\\
n'_\mu&=&{\Delta\over 2\Sigma}\left(\delta_\mu^t-a\sin^2\vartheta \delta_\mu^\varphi\right)
+{1\over 2}\delta_\mu^r, \nonumber\\
m_\mu&=&
{i\sin\vartheta\over 2^{1/2}(r+ia\cos\vartheta)}\left[
a\delta_\mu^t-(r^2+a^2)\delta_\mu^\varphi\right]
-{r-ia\cos\vartheta\over 2^{1/2}}\delta_\mu^\vartheta, \nonumber\\
m'_\mu&=&\bar m_\mu. \label{eq:null_tetrad}
\end{eqnarray}
where the bar denotes the complex conjugation.
These are subject to the normalization: $n_\mu n'^\mu=-m_\mu m'^\mu=1$,
$n_\mu n^\mu=n'_\mu n'^\mu=m_\mu m^\mu=n_\mu m^\mu=n'_\mu m^\mu=0$.
Alternative description is given by the two-component spinor $o^A,\iota^A$ via
the identification
\begin{eqnarray}
(n^\mu,n'^\mu,m^\mu,m'^\mu)
\leftrightarrow(o^A\bar o^{A'},\iota^A\bar\iota^{A'},o^A\bar\iota^{A'},\iota^A\bar o^{A'}),
\end{eqnarray}
with the symplectic structure $\epsilon^{AB}=o^A\iota^B-\iota^A o^B$,
$\epsilon^{01}=\epsilon_{01}=1$.
Each component of the spinor covariant derivative $\nabla_{AA'}$ is denoted by
\begin{equation}
(\nabla_{o\bar o},
\nabla_{\iota\bar \iota},
\nabla_{o\bar \iota},
\nabla_{\iota\bar o})
=(D,D',\delta,\delta'),
\end{equation}
and the spin-coefficients are defined by
\begin{eqnarray}
D(o,\iota)&=&(\epsilon o-\kappa\iota,-\tau' o-\epsilon\iota),~~~
D'(o,\iota)=(-\epsilon' o-\tau\iota,-\kappa'o+\epsilon'\iota),\nonumber\\
\delta(o,\iota)&=&(\beta o-\sigma\iota,-\rho'o-\beta\iota),~~~
\delta'(o,\iota)=(-\beta'o-\rho\iota,-\sigma'o+\beta'\iota).\nonumber\\
\end{eqnarray}
Then, all the nonzero spin-coefficients are
\footnote{
Though we have defined the
spin-coefficients in spinor form,
the tensor calculus would work better in actual computation.
See eqs.~(4.5.28) in ref.~\cite{Penrose:1985jw} for the equivalent
tensorial definition
for the spin-coefficient.
}
\begin{eqnarray}
\tau&=&-{ia\sin\vartheta\over 2^{1/2}\Sigma},~~~
\rho=-{1\over r-ia\cos\vartheta},~~~
\beta=-{\bar\rho\cot\vartheta\over2\sqrt{2}},~~~
\tau'=-{ia\rho^2\sin\vartheta\over \sqrt{2}},\nonumber\\
\rho'&=&-{\rho^2\bar\rho\Delta\over 2},~~~
\epsilon'=\rho'-{\rho\bar\rho\over 4}\Delta_{,r},~~~
\beta'=\tau'+\bar\beta.
\end{eqnarray}

Here, let us consider the Weyl equation ($s=1/2$) and the Maxwell equation ($s=1$) on
this background brane metric.

We define the component of the Weyl spinor $\psi_A$ simply by
 $\psi_0=\psi_Ao^A$ and $\psi_1=\psi_A\iota^A$.
Then, each component of the Weyl equation $\nabla^{AA'}\psi_A$ becomes explicitly
\begin{eqnarray}
D\psi_1-\delta'\psi_0&=&(\beta'-\tau')\psi_0+(\rho-\epsilon)\psi_1,
\label{Weyl-1}\\
\delta\psi_1-D'\psi_0&=&(\epsilon'-\rho')\psi_0+(\tau-\beta)\psi_1.
\label{Weyl-2}
\end{eqnarray}

On the other hand, the Maxwell field is represented by the
second-rank symmetric spinor $\phi_{AB}$, and its components are
denoted by $\phi_0=\phi_{AB} o^Ao^B$, $\phi_1=\phi_{AB}
o^A\iota^B$ and $\phi_2=\phi_{AB} \iota^A\iota^B$, respectively.
Then, the source-free Maxwell equation $\nabla^{AA'}\phi_{AB}=0$
leads to
\begin{eqnarray}
D\phi_1-\delta'\phi_0&=&(2\beta'-\tau')\phi_0+2\rho\phi_1-\kappa\phi_2
\label{Maxwell-1}\\
D\phi_2-\delta'\phi_1&=&\sigma'\phi_0-2\tau'\phi_1+(\rho-2\epsilon)\phi_2
\label{Maxwell-2}\\
D'\phi_0-\delta\phi_1&=&(\rho'-2\epsilon')\phi_0-2\tau\phi_1+\sigma\phi_2
\label{Maxwell-3}\\
D'\phi_1-\delta\phi_2&=&-\kappa'\phi_0+2\rho'\phi_1+(2\beta-\tau)\phi_2.
\label{Maxwell-4}
\end{eqnarray}

The brane-induced metric turns out to be of Petrov type D,
namely, the gravi\-ta\-tion\-al spinor $\Psi_{ABCD}$ has only nonzero component
$\Psi_2$$=$$\Psi_{ABCD}o^Ao^B\iota^C\iota^D$.
Besides this condition, when $\kappa=\sigma=\kappa'=\sigma'=0$ hold as in the present case,
we have the identities for the differential operators
\begin{eqnarray}
&&[D-(p+1)\epsilon+\bar\epsilon+q\rho-\bar\rho](\delta-p\beta+q\tau)
\nonumber\\
&&{}
-[\delta-(p+1)\beta+\bar\beta'-\bar\tau'+q\tau](D-p\epsilon+q\rho)
=0,
\label{commute}\\
&&[D'-(p+1)\epsilon'+\bar\epsilon'+q\rho'-\bar\rho'](\delta'-p\beta'+q\tau')
\nonumber\\
&&{}-[\delta'-(p+1)\beta'+\bar\beta-\bar\tau+q\tau'](D'-p\epsilon'+q\rho')
=0,
\label{commute'}
\end{eqnarray}
for any pair of the numbers ($p,q$), where we have used the identities
\begin{eqnarray}
\delta D-D\delta&=&(\bar{\tau'}-\bar{\beta'}+\beta)D+\kappa D'
-(\bar{\rho}+\epsilon-\bar{\epsilon})\delta-\sigma\delta',\\
\delta' D'-D'\delta'&=&\kappa' D+(\bar{\tau}-\bar{\beta}+\beta')D'
-\sigma'\delta-(\bar{\rho'}+\epsilon'-\bar{\epsilon'})\delta'.
\end{eqnarray}
Applying $(\delta+\bar\beta'-\bar\tau'-\tau)$ to Eq.~(\ref{Weyl-1})
and $(D+\bar\epsilon-\rho-\bar\rho)$ to Eq.~(\ref{Weyl-2}),  subtracting one from
the other,
and using Eq.~(\ref{commute}) for $(p,q)=(-1,-1)$,
we obtain the decoupled equation for $\psi_0$
\begin{eqnarray}
&&\Biggl\{
\left[{(r^2+a^2)^2\over\Delta}-a^2\sin^2\vartheta\right]
{\partial^2\over\partial t^2}
+2a\left({r^2+a^2\over\Delta}-1\right)
{\partial^2\over\partial t\partial\varphi}
+\left({a^2\over\Delta}-{1\over\sin^2\vartheta}\right)
{\partial^2\over\partial \varphi^2}\nonumber\\
&&{}
+\left[
2r-{(r^2+a^2)\Delta_{,r}\over2\Delta}
+ia\cos\vartheta
\right]
{\partial\over\partial t}
+\left[
-{a\Delta_{,r}\over2\Delta}-{i\cos\vartheta\over\sin^2\vartheta}
\right]
{\partial\over\partial \varphi}
\nonumber\\
&&{}
-\Delta^{-1/2}{\partial\over\partial r}\Delta^{3/2}{\partial\over\partial r}
-{1\over\sin\vartheta}{\partial\over\partial\vartheta}
\sin\vartheta{\partial\over\partial\vartheta}
+{\cot^2\vartheta\over 4}-{1\over 2}+{n(n-1)\mu r^{-n-1}\over 2}
\Biggr\}\psi_0=0.\nonumber\\
\end{eqnarray}
If we set $\psi_0=R(r)S(\vartheta)e^{-i\omega t+i m\varphi}$, then we obtain
\begin{eqnarray}
 &&{1\over\sin\vartheta}{d\over d\vartheta}\left(\sin\vartheta
{d S\over d\vartheta}\right)\nonumber\\
&&{}+\left(
a^2\omega^2\cos^2\vartheta-{m^2\over\sin^2\vartheta}-a\omega\cos\vartheta
-{m\cos\vartheta\over\sin^2\vartheta}-{1\over 4}\cot^2\vartheta+{1\over 2}
+A
\right)S=0,\nonumber\\\\
&&\Delta^{-1/2}{d\over dr}\left(\Delta^{3/2}{dR\over dr}\right)\nonumber\\
&&{}+\Biggl[
\left(
{K^2\over\Delta}
+2i\omega r
-{i\over2}
{[2r+(n-1)\mu r^{-n}]K\over\Delta}
\right)\nonumber\\
&&
{}
-{n(n-1)\mu r^{-n-1}\over 2}
+2ma\omega-a^2\omega^2-A
\Biggl]R=0.
\end{eqnarray}

For the Maxwell field,
applying $(\delta-\beta+\bar\beta'-\bar\tau'-2\tau)$ to Eq.~(\ref{Maxwell-1})
and
$(D-\epsilon+\bar\epsilon-2\rho-\bar\rho)$ to Eq.~(\ref{Maxwell-2}),
subtracting one from the other,
and using Eq.~(\ref{commute}) for $(p,q)=(0,-2)$, we obtain
\begin{eqnarray}
&&\Biggl\{
\left[
{(r^2+a^2)^2\over\Delta}
-a^2\sin^2\vartheta\right]
{\partial^2\over\partial t^2}
+\left[{2a(r^2+a^2)\over \Delta}
-2a\right]
{\partial^2\over\partial t\partial\varphi}
+\left[{a^2\over \Delta}-{1\over\sin^2\vartheta}\right]
{\partial^2\over\partial \varphi^2}
\nonumber\\
&&
{}+\left[-{\mu r^{-n}\left[(n+1)r^2+(n-1)a^2\right]\over\Delta}
+2(r+ia\cos\vartheta)\right]
{\partial\over\partial t}
\nonumber\\
&&{}+\left[
-{a[2r+(n-1)\mu r^{-n}]\over\Delta}-{2i\cos\vartheta\over\sin^2\vartheta}
\right]
{\partial\over\partial \varphi}
\nonumber\\
&&
{}-{1\over\Delta}{\partial\over\partial r}\Delta^2{\partial\over\partial r}
-{1\over\sin\vartheta}{\partial\over\partial \vartheta}
\sin\vartheta{\partial\over\partial \vartheta}
+\cot^2\vartheta-1+n(n-1)\mu r^{-n-1} \Biggr\}\varphi_0=0. \nonumber\\
\end{eqnarray}
Set $\phi_0=R(r)S (\vartheta)e^{-i\omega t+im\varphi}$, then we
have
\begin{eqnarray}
  &&{1\over\sin\vartheta}{d\over d\vartheta}\left(\sin\vartheta
{d S\over d\vartheta}\right)\nonumber\\
&&{}+\left(
a^2\omega^2\cos^2\vartheta-{m^2\over\sin^2\vartheta}-2a\omega\cos\vartheta
-{2m\cos\vartheta\over\sin^2\vartheta}-\cot^2\vartheta+1
+A
\right)S=0,\nonumber\\\\
&&{1\over\Delta}{d\over dr}\left(\Delta^2{dR\over dr}\right)\nonumber\\
&&{}+\Biggl[
{K^2\over\Delta}
+4i\omega r-i
{[2r+(n-1)\mu r^{-n}]K\over\Delta}
\nonumber\\
&&
{}
-n(n-1)\mu r^{-n-1}
+2ma\omega-a^2\omega^2-A
\Biggl]R=0.
\end{eqnarray}

In summary, the spin-$s$ massless field equation becomes
\begin{eqnarray}
&&{1\over\sin\vartheta}{d\over d\vartheta}\left(\sin\vartheta
{d S\over d\vartheta}\right)
\nonumber\\
&&{}
+\left[
(s-a\omega\cos\vartheta)^2
-
(s\cot\vartheta+m\csc\vartheta)^2
-s(s-1)+A
\right]S=0,\nonumber\\
\end{eqnarray}
and
\begin{eqnarray}
&&\Delta^{-s}{d\over dr}\left(\Delta^{s+1}{dR\over dr}\right)+\Biggl[
{K^2\over\Delta}
+s\left(
4i\omega r
-i{[2r+(n-1)\mu r^{-n}]K\over\Delta}-n(n-1)\mu r^{-n-1}\right)
\nonumber\\
&&{}
+2ma\omega-a^2\omega^2-A
\Biggl]R=0.
\end{eqnarray}

\bibliography{paper}
\bibliographystyle{utphys}

\newpage

\end{document}